\documentclass[showpacs,showkeys,preprintnumbers,amsmath,amssymb]{revtex4-1}
\usepackage{graphicx}
\usepackage{dcolumn}
\usepackage{float}
\usepackage{url}
\usepackage{siunitx}
\usepackage{amsfonts}
\usepackage{color}

\newcommand{\abs}[1]{\left| #1 \right|}
\newcommand{\bd}{\partial}
\newcommand{\arccot}{\operatorname{arccot}}
\renewcommand{\Re}{\operatorname{Re}}
\renewcommand{\Im}{\operatorname{Im}}

\begin{document}

\title{Squeezing quadrature rotation in the acoustic band via optomechanics}
\author{Giovanni Guccione}
\email{giovanni.guccione@anu.edu.au}
\author{Harry J.\ Slatyer}
\author{Andr\'e R.\ R.\ Carvalho}
\author{Ben C.\ Buchler}
\author{Ping Koy Lam}
\email{Ping.Lam@anu.edu.au}
\affiliation{Centre for Quantum Computation and Communication Technology, Department of Quantum Science, The Australian National University, Canberra, Australia}
\date{\today}
\begin{abstract}
We examine the use of optomechanically-generated squeezing to obtain a sensitivity enhancement for interferometers in the gravitational-wave band. The intrinsic dispersion characteristics of optomechanical squeezing around the mechanical frequency are able to produce squeezing at different quadratures over the spectrum, a feature required by gravitational-wave interferometers to beat the standard quantum limit over an extended frequency range. Under realistic assumptions we show that the amount of available squeezing and the intrinsic quadrature rotation may provide, compared to similar amounts of fixed-quadrature squeezing, a detection advantage. A significant challenge for this scheme, however, is the amount of excess noise that is generated in the unsqueezed quadrature at frequencies near the mechanical resonance.
\end{abstract}

\maketitle

\section{Introduction}
\label{sec: Introduction}

As technology advancements push precision metrology to new sensitivity limits, the complications and trade-offs arising from quantum effects become increasingly significant. Heisenberg's uncertainty principle encapsulates the effects of the quantum noise that is the ultimate limit to measurements. For optical measurements made with an ideal laser source, the quantum noise manifests as equal uncertainty in both the amplitude and phase quadratures of the light. This is the physical origin of photon shot noise~\cite{Yuen:1983:OptLett}. In a precision measurement where all sources of technical noise have been mitigated so that only the shot noise of the light remains, the signal-to-noise ratio can be enhanced by increasing the power of the laser. In this way the impact of shot noise can be reduced, but not indefinitely: the resources available may be limited, and excessive amounts of power may start introducing additional sources of noise. Even under optimal conditions for the control of thermal effects, the radiation pressure of the intense field inevitably impinges on the measurement system. As the shot noise performance is improved by increasing power, the quantum radiation pressure noises places a new limit on the measurement. The trade-off between shot noise and radiation pressure noise is a manifestation of the ``Standard Quantum Limit" (SQL)~\cite{Braginsky:1968:JExpTheorPhys,Caves:1980:RevModPhys}. It is this issue that will pose a problem to long-baseline interferometric gravitational-wave sensors~\cite{Caves:1981:PhysRevD} in the near future. The SQL can be broken, but one requires modification of the uncertainty in the quadratures of the optical field, i.e.\ a squeezed state of light~\cite{Kimble:2001:PhysRevD}.

The squeezing of coherent states of light has seen many breakthroughs in recent years, with great improvements in efficiency and robustness. Established techniques based on non-linearities from optical parametric oscillators have accomplished up to \SI{12.7}{\decibel} of noise cancellation below the shot~noise~level~\cite{Eberle:2010:PhysRevLett}, and novel methods for generating squeezing keep attracting the interest of communities operating in regimes at different frequency bands or requiring frequency-dependence properties.

In regard to gravitational-wave interferometers, squeezed light is already in use to push the sensitivity beyond the standard quantum limit (SQL)~\cite{Vahlbruch:2006:PhysRevLett, Aasi:2013:NatPhot}. The current procedure employs phase-squeezed light, which is effective against the photon shot noise at higher frequencies. This is optimal provided the laser power is not high enough to cause radiation pressure noise at low frequencies ~\cite{Caves:1981:PhysRevD}. The next generation of gravitational wave detectors will have high enough laser power to generate significant quantum radiation pressure at low frequencies, and their sensitivity would be enhanced by a phase-squeezed state due to excess noise of its amplitude quadrature. It is well known that quantum radiation pressure noise and photon shot noise can be simultaneously suppressed if the squeezing angle is frequency dependant, with amplitude squeezing to reduce radiation pressure noise at low frequencies and phase squeezing at higher frequencies to reduce shot noise ~\cite{Unruh:1982:Springer, Kimble:2001:PhysRevD, Chelkowski:2005:PhysRevA}. Broadband enhancement would thus be achieved when the squeezed angle rotates by $\pi/2$ over frequencies around the point of optimal sensitivity for gravitational waves, which is approximately $\SI{100}{\hertz}$.

The dispersive properties of filter cavities can achieve the desired quadrature rotation starting from a fixed-angle squeezing source~\cite{Kimble:2001:PhysRevD, Harms:2003:PhysRevD, Khalili:2010:PhysRevD, Evans:2013:PhysRevD}, and the idea has already been implemented with proof-of-principle demonstrations~\cite{Chelkowski:2005:PhysRevA,Oelker:2015:arXiv}. Technical impediments such as decoherence and degradation can impact the effectiveness of this method~\cite{Kwee:2014:PhysRevD}, and cavity losses represent the most significant limitation: to achieve the necessary storage time, the length of the resonators would need to be between a few tens of meters and the entire length of the interferometer arms~\cite{Corbitt:2008:PhD}. Optomechanically-induced transparency (OMIT)~\cite{Weis:2010:Science} could implement dispersion over a narrower bandwidth~\cite{Ma:2014:PhysRevLett, Qin:2014:PhysRevA} and qualifies as a suitable candidate to achieve frequency-dependent noise cancellation in the right frequency band. The same principle has inspired other proposals, such as the inclusion of a feedback-controlled unstable optomechanical system within the signal-recycling cavities~\cite{Miao:2015:PhysRevLett}.

We aim to extend the set of tools available to gravitational-wave detectors by exploring the effects of injecting squeezing generated by an optomechanical system~\cite{Fabre:1994:PhysRevA, Mancini:1994:PhysRevA, Heidmann:1994:PhysRevA, Nunnenkamp:2010:PhysRevA} into the system. Quadrature rotation naturally occurs in optomechanical squeezing due to the dispersive nature of the interaction between light and mechanics. Appropriate tuning of the optomechanical cavity could make this rotation proximate to the one required for ideal interferometric measurements, providing an alternative to filter cavities and fixed-angle squeezing injection. Recent experimental demonstrations~\cite{Brooks:2012:Nature, Safavi-Naeini:2013:Nature, Purdy:2013:PhysRevX} show that optomechanical squeezing is rapidly growing more enticing for applications beyond proof-of-principle, and suggest that implementation of our proposal could be within reach of a present state-of-the-art system.

\section{Optomechanical Squeezing}
\label{sec: Optomechanical Squeezing}

The crucial element for the generation of squeezed states is a source of field non-linearity that introduces cross-correlations between different noise quadratures, opening up the opportunity to transfer the uncertainty of one onto the other. Typically, the non-linear optical interaction is obtained through optical parametric oscillator (OPO) crystals, whose refractive index depends on the intensity of the field traversing the material. Cavity optomechanics presents an alternative method to generate analogous non-linear effects thanks to the mutual dependence of the intensity of the intra-cavity field and the relative position of the end mirrors of the cavity.

Assuming a linear optical cavity with one fixed input mirror and the other end mirror oscillating at a mechanical frequency $\omega_\textrm{m}$, the Hamiltonian describing the system is
\begin{equation}
	\hat{H} = \hbar\omega(\hat{x})\Big(\hat{a}^\dag\hat{a}+\frac{1}{2}\Big) + \frac{\hat{p}^2}{2m} + \frac{1}{2}m\omega_\textrm{m}^2\hat{x}^2,
\label{eqn: hamiltonian}
\end{equation}
where $\hat{a}$ and $\hat{a}^\dag$ are the operators of the optical mode within the cavity, $\hat{x}$ and $\hat{p}$ are the position and momentum operators of the moving mirror and $m$ is its mass. The resonance frequency of the cavity $\omega(\hat{x})$ can be expanded in terms of the mirror position to make the optomechanical interaction explicit. To linear orders in $\hat{x}$, ignoring the constant terms and the contribution of the vacuum to the interaction, the Hamiltonian is
\begin{equation}
	\hat{H} \simeq \hbar\omega_\textrm{c}\hat{a}^\dag\hat{a} + \frac{\hat{p}^2}{2m} + \frac{1}{2}m\omega_\textrm{m}^2\hat{x}^2 - \hbar G_0\hat{a}^\dag\hat{a}\hat{x},
\label{eqn: expanded hamiltonian}
\end{equation}
where $\omega_\textrm{c}$ is the cavity frequency when the mirror is in the rest position and $G_0:=-\bd\omega(x)/\bd x$ is the optomechanical coupling strength, equal to $\omega_\textrm{c}/L$ for a linear cavity of length $L$. For the full dynamics of the system we need to include fluctuation-dissipation terms due to the coupling of both optical and mechanical degrees of freedom with external baths. The mechanical oscillations are intrinsically excited by the Brownian forces $\hat{F}_\textrm{th}$ due to the thermal bath, to which the oscillator is coupled by means of the mechanical damping rate $\gamma$. The damping rate is related to the mechanical frequency via the mechanical quality factor $Q_\textrm{m}:=\omega_\textrm{m}/\gamma$. The stochastic thermal forces are taken to be Hermitian and to have zero expectation value~\cite{Gardiner:2004:Springer}. The optical mode is driven by the input field $\hat{a}_\textrm{in}$, whose expectation value is given by the complex field amplitude $\alpha_\mathrm{in}$, and the dissipation due to the optical losses is expressed in terms of the cavity half-linewidth $\kappa$. Moving to a frame rotating at the optical frequency $\omega_\textrm{o}=2\pi c/\lambda$, and introducing the preliminary cavity detuning $\Delta_0 = \omega_\textrm{o}-\omega_\textrm{c}$, the equations of motion for $\hat{x}$ and $\hat{a}$ are
\begin{equation}
	\begin{cases}
		m\ddot{\hat{x}} = -m\omega_\mathrm{m}^2\hat{x}+\hbar G_0\hat{a}^\dag\hat{a}-m\gamma\dot{\hat{x}}+\hat{F}_\textrm{th},	\\[0.8em]
		\dot{\hat{a}} = i(\Delta_0+G_0\hat{x})\hat{a}-\kappa\hat{a}+\sqrt{2\kappa}\hat{a}_\mathrm{in}.
	\end{cases}
\label{eqn: equations of motion}
\end{equation}
The coefficient $G_0$ mediates the coupling between the optical and the mechanical modes: the position of the mirror is affected by an intensity-dependent term representing the effect of radiation pressure force; at the same time, the cavity field experiences a position-dependent phase shift which corresponds to the change of resonance frequency due to the modified cavity length. In its present form the equations of motions are non-linear, and before looking at the quadratures of the optical field we will need to transform the variables and consider their fluctuation terms. It is still of interest to consider Eq.~\ref{eqn: equations of motion} for solutions of the expected values in the steady state:
\begin{equation}
	\begin{cases}
		x_\textrm{s} = \frac{\hbar G_0\abs{\alpha_s}^2}{m\omega_\textrm{m}^2},	\\[0.8em]
		\alpha_\textrm{s} = \frac{\sqrt{2\kappa}\alpha_\textrm{in}}{\kappa-i(\Delta_0+G_0x_\textrm{s})}.
	\end{cases}
\label{eqn: steady state solutions}
\end{equation}
Combining the identities of Eq.~\ref{eqn: steady state solutions} into the cubic relation for $x_\textrm{s}$ (or equivalently $\abs{\alpha_\textrm{s}}^2$) we get evidence of the optomechanical bistability:
\begin{equation}
	x_\textrm{s}[\kappa^2+(\Delta_0+G_0x_\textrm{s})^2] = \frac{\hbar G_0}{m\omega_\textrm{m}^2}2\kappa\abs{\alpha_\textrm{in}}^2,
\label{eqn: bistability condition}
\end{equation}
A direct consequence of Eq.~\ref{eqn: bistability condition} is that, for a parameter regime which allows the quadratic $3G_0^2x_\textrm{s}^2+4\Delta_0G_0x_\textrm{s}+\kappa^2+\Delta_0^2$ to attain negative values, more than one solution is possible. This arises from the fact that for high coupling strengths (or high powers) the thrust of radiation pressure force pushes the mirror enough to detune the cavity outside of its linewidth, even during the build-up of the intra-cavity field. The asymmetry of the ponderomotive detuning skews the common intensity profile of the cavity response, creating a region where two stable (and one unstable) solutions are possible.

To look at the dynamics of small fluctuations we consider perturbation of the variables around their corresponding steady states:
\begin{align*}
	\hat{x} \rightarrow\,	&	x_\textrm{s}+\delta\hat{x},		&	\hat{F}_\textrm{th} \rightarrow\,	&	\delta\hat{F}_\textrm{th},	\\
	\hat{a} \rightarrow\,	&	\alpha_\textrm{s}+\delta\hat{a},	&	\hat{a}_\textrm{in} \rightarrow\,			&	\alpha_\textrm{in}+\delta\hat{a}_\textrm{in}.
\label{eqn: perturbed operators}
\end{align*}
The non-vanishing correlations of the optical input modes, $\langle\delta\hat{a}_\textrm{in}^\dag(\omega)\delta\hat{a}_\textrm{in}(\omega')\rangle=2\pi\delta(\omega+\omega')n_\textrm{o}^\textrm{th}$ and $\langle\delta\hat{a}_\textrm{in}(\omega)\delta\hat{a}_\textrm{in}^\dag(\omega')\rangle = 2\pi\delta(\omega+\omega')(1+n_\textrm{o}^\textrm{th})$, increase with the mean thermal occupation of the photons $n_\textrm{o}^\textrm{th}$. As we consider coherent states satisfying $\hbar\omega_\textrm{o}\gg k_\textrm{B}T$, we can validly approximate $n_\textrm{o}^\textrm{th}\simeq0$. For the thermal drive of the mechanical oscillator we assume that the system is in the fast thermal correlation time limit~\cite{Gardiner:2004:Springer, Giovannetti:2001:PhysRevA} and require the stochastic force term to be Markovian: $\langle\delta\hat{\epsilon}_\textrm{th}(\omega)\delta\hat{\epsilon}_\textrm{th}(\omega')\rangle = 2\pi\delta(\omega+\omega')S_{F}^\textrm{th}(\omega)$. The thermal spectrum $S_{F}^\textrm{th}(\omega)=m\gamma\hbar\omega(2n_\textrm{m}^\textrm{th}+1)$ can be approximated to $S_{F}^\textrm{th}(\omega)\simeq2m\gamma k_\textrm{B}T$ when the thermal phonon occupation number is $n_\textrm{m}^\textrm{th}\gg1$ at high temperatures. Note that the main noise contribution for the system, coming from thermal fluctuations, is directly proportional to the temperature but inversely proportional to the mechanical quality factor. In terms of the thermal noise, having the oscillator at a specific temperature is equivalent to having a better oscillator at higher temperature, and it is ultimately the ratio $T/Q_\textrm{m}$ that determines the coherence of the system and the squeezing attainable. Even though extremely high mechanical quality factors have been reported at room temperature~\cite{Weaver:2015:arXiv,Reinhardt:2015:arXiv}, lower temperatures are still appealing as they allow a relaxation of the requirements on the oscillator.

After the expansion in terms of the fluctuation variables, we can identify two main interaction terms: one is linear in $\delta\hat{a}$ and proportional to the complex field amplitude $\alpha_\textrm{s}$, the other is non-linear and quadratic in the fluctuations of the optical mode. Allowing $\abs{\alpha_\textrm{s}}\gg1$, the non-linear interaction has a negligible contribution and one obtains linearized equations of motion:
\begin{equation}
	\begin{cases}
		m\delta\ddot{\hat{x}} = -m\omega_\mathrm{m}^2\delta\hat{x}+\hbar G_0(\alpha_\textrm{s}\delta\hat{a}^\dag + \alpha_\textrm{s}^*\delta\hat{a})-m\gamma\delta\dot{\hat{x}} + \delta\hat{F}_\textrm{th},	\\[0.8em]
		\delta\dot{\hat{a}} = (-\kappa+i\Delta)\delta\hat{a}+iG_0\alpha_\textrm{s}\delta\hat{x}+\sqrt{2\kappa}\delta\hat{a}_\mathrm{in},
	\end{cases}
\label{eqn: linearized equations of motion}
\end{equation}
where we introduced the effective detuning $\Delta = \Delta_0 + G_0x_\textrm{s}$. To solve the linear system of differential equations we move to the Fourier domain with the transform $\tilde{f}(\omega):=\int_{-\infty}^{+\infty}f(t)e^{-i\omega t}dt$ (the tilded superscript will be dropped for simplicity of notation). Once the solution for $\delta\hat{a}$ is found, we use the input-output relation $\delta\hat{a}_\textrm{out}=\sqrt{2\kappa}\delta\hat{a}-\delta\hat{a}_\textrm{in}$ to obtain an explicit expression for the output field. We define the original mechanical susceptivity $\chi_\textrm{m}(\omega):=[m(\omega_\textrm{m}^2-\omega^2-i\gamma\omega)]^{-1}$ and the effective susceptivity $\chi_\textrm{eff}(\omega):=[\chi_\textrm{m}(\omega)^{-1}+i\hbar G_0^2\abs{\alpha_\textrm{s}}^2(\mathcal{A}_{-}(\omega)-\mathcal{A}_{+}(\omega))]^{-1}$, which accounts for the correction due to the optical spring effect and depends on the Airy functions $\mathcal{A}_{-}(\omega):=[\kappa+i(\Delta-\omega)]^{-1}$ and $\mathcal{A}_{+}(\omega):=[\kappa-i(\Delta+\omega)]^{-1}$. 
The solution for the output field is then expressed as
\begin{eqnarray}
	\begin{split}
		\delta\hat{a}_\textrm{out}(\omega) =	&	\Big(i\hbar G_0^2\abs{\alpha_\textrm{s}}^2(2\kappa)\mathcal{A}_{-}(\omega)\mathcal{A}_{+}(\omega)\chi_\textrm{eff}(\omega)\\
		& \phantom{\bigg(}+(2\kappa)\frac{\chi_\textrm{eff}(\omega)}{\chi_\textrm{m}(\omega)}\mathcal{A}_{+}(\omega)-1\Big)\,\delta\hat{a}_\textrm{in}	\\
		& + i\hbar G_0^2\alpha_\textrm{s}^2(2\kappa)\mathcal{A}_{-}(\omega)\mathcal{A}_{+}(\omega)\chi_\textrm{eff}(\omega)\;\delta\hat{a}_\textrm{in}^\dag	\\
		& + iG_0\alpha_\textrm{s}\sqrt{2\kappa}\mathcal{A}_{+}(\omega)\chi_\textrm{eff}(\omega)\;\delta\hat{F}_\textrm{th}.
	\end{split}
\label{eqn: output field solution}
\end{eqnarray}
Remembering that $[\tilde{f}(\omega)]^\dag=\tilde{f^\dag}(-\omega)$, one can obtain a similar expression for $\delta\hat{a}_\textrm{out}^\dag(\omega)$. Differently from the case of a purely optical cavity (easily obtained in the limit of $G_0\rightarrow0$), the quantum fluctuation operators of the output field $\delta\hat{a}$ and $\delta\hat{a}^\dag$ are correlated with each other. This arises from the fact that the original input shot noise acts on the mirror which then transduces the fluctuations back to the optical field. Entering from two channels, the noise can destructively interfere and the optical modes can undergo a reduction in quantum fluctuations, or squeezing. It should be noted that a similar response is obtained by dissipative optomechanics~\cite{Qu:2015:PhysRevA,Kilda:2016:JOpt}, however the typical coupling strengths of this scheme are generally too weak to apply a significant contribution and will not be considered for the following analysis. From Eq.~\ref{eqn: output field solution} one can recognize that the effective mechanical susceptivity $\chi_\textrm{eff}(\omega)$ plays a role analogous to that of the optomechanical coupling strength $G_0$ in the cross-coupling. This factor, opportunely modulated by the Airy functions $\mathcal{A}_{\pm}(\omega)$, contributes to the frequency dependence of the correlations terms at the origin of optomechanical squeezing. It is also a consequence of $\chi_\textrm{eff}(\omega)$ that the coupled dynamics vanish at high frequencies.

Looking now at the quadratures of the output field, which we describe in a generic parametric form as $\hat{X}_{\theta}^\textrm{out}:=e^{i\theta}\delta\hat{a}_\textrm{out}+e^{-i\theta}\delta\hat{a}_\textrm{out}^\dag$, we wish to get an expression for the spectral density, normalized to the shot noise, $S_{\theta}(\omega):=\int_{-\infty}^{+\infty}\frac{d\omega'}{2\pi}\langle\{\hat{X}_{\theta}^\textrm{out}(\omega),\hat{X}_{\theta}^\textrm{out}(\omega')\}\rangle$. The curly brackets indicate symmetrization over the variables: $\{\tilde{f}(\omega),\tilde{g}(\omega')\}:=(\tilde{f}(\omega)\tilde{g}(\omega')+\tilde{f}(\omega')\tilde{g}(\omega))/2$.
Considering two orthogonal quadratures as reference, the parametric spectrum can be expanded into a more convenient form~\cite{Collett:1985:PhysRevA} as
\begin{equation}
	\begin{split}
		S_{\theta}(\omega) =	&	\frac{S_{X}(\omega)+S_{Y}(\omega)}{2} + \frac{S_{X}(\omega)-S_{Y}(\omega)}{2}\cos{(2\theta)}	\\
						&	+ S_{XY}(\omega)\sin{(2\theta)},
\label{eqn: parametric spectrum}
	\end{split}
\end{equation}
where $S_{X}(\omega):=S_{0}(\omega)$ and $S_{Y}(\omega):=S_{\pi/2}(\omega)$ are the spectral densities of the two reference quadratures and $S_{XY}(\omega):=\frac{1}{2}\int_{-\infty}^{+\infty}\frac{d\omega'}{2\pi}\langle\{\hat{X}_\textrm{out}^{0}(\omega),\hat{X}_\textrm{out}^{\pi/2}(\omega')\}+\{\hat{X}_\textrm{out}^{\pi/2}(\omega),\hat{X}_\textrm{out}^{0}(\omega')\}\rangle$ is the spectral density of their cross-correlation. Explicit formulations for each element of Eq.~\ref{eqn: parametric spectrum} can be obtained by plugging the solution of the output field operator (Eq.~\ref{eqn: output field solution}) into the expression of the quadrature of interest and expanding the non-vanishing correlation terms for $\delta\hat{a}_\textrm{in}$ and $\delta\hat{F}_\textrm{th}$. This results in:
\begin{widetext}
\begin{equation}
	\begin{cases}
			S_{X}(\omega) = \frac{1}{2}\Big(\abs{-i\hbar G_0^2(\abs{\alpha_\textrm{s}}^2-\alpha_\textrm{s}^2)(2\kappa)\mathcal{A}_{-}(\omega)\mathcal{A}_{+}(\omega)\chi_\textrm{eff}(\omega)+(2\kappa)\mathcal{A}_{-}(\omega)\frac{\chi_\textrm{eff}(\omega)}{\chi_\textrm{m}(\omega)}-1}^2	\\
			\phantom{S_{X}(\omega) = \frac{1}{2}\Big(}+\abs{i\hbar G_0^2(\abs{\alpha_\textrm{s}}^2-{\alpha_\textrm{s}^*}^2)(2\kappa)\mathcal{A}_{-}(\omega)\mathcal{A}_{+}(\omega)\chi_\textrm{eff}(\omega)+(2\kappa)\mathcal{A}_{+}(\omega)\frac{\chi_\textrm{eff}(\omega)}{\chi_\textrm{m}(\omega)}-1}^2	\Big)\\
			\phantom{S_{X}(\omega) = }+(2\kappa)\abs{\chi_\textrm{eff}(\omega)}^2\abs{G_0\alpha_\textrm{s}\mathcal{A}_{+}(\omega)-G_0\alpha_\textrm{s}^*\mathcal{A}_{-}(\omega)}^2\,S_{F}^\textrm{th}(\omega),	\\[0.8em]
			S_{Y}(\omega) = \frac{1}{2}\Big(\abs{-i\hbar G_0^2(\abs{\alpha_\textrm{s}}^2+\alpha_\textrm{s}^2)(2\kappa)\mathcal{A}_{-}(\omega)\mathcal{A}_{+}(\omega)\chi_\textrm{eff}(\omega)+(2\kappa)\mathcal{A}_{-}(\omega)\frac{\chi_\textrm{eff}(\omega)}{\chi_\textrm{m}(\omega)}-1}^2	\\
			\phantom{S_{Y}(\omega) = \frac{1}{2}\Big(}+\abs{i\hbar G_0^2(\abs{\alpha_\textrm{s}}^2+{\alpha_\textrm{s}^*}^2)(2\kappa)\mathcal{A}_{-}(\omega)\mathcal{A}_{+}(\omega)\chi_\textrm{eff}(\omega)+(2\kappa)\mathcal{A}_{+}(\omega)\frac{\chi_\textrm{eff}(\omega)}{\chi_\textrm{m}(\omega)}-1}^2	\Big)\\
			\phantom{S_{Y}(\omega) = }+(2\kappa)\abs{\chi_\textrm{eff}(\omega)}^2\abs{G_0\alpha_\textrm{s}\mathcal{A}_{+}(\omega)+G_0\alpha_\textrm{s}^*\mathcal{A}_{-}(\omega)}^2\,S_{F}^\textrm{th}(\omega),	\\[0.8em]
			S_{XY}(\omega) = \Im\!\Big[\big(-i\hbar G_0^2\abs{\alpha_\textrm{s}}^2(2\kappa)\mathcal{A}_{-}(\omega)\mathcal{A}_{+}(\omega)\chi_\textrm{eff}(\omega)+(2\kappa)\mathcal{A}_{-}(\omega)\frac{\chi_\textrm{eff}(\omega)}{\chi_\textrm{m}(\omega)}-1\big)	\\
			\phantom{S_{XY}(\omega) = \Im\!\Big[}\times\big(i\hbar G_0^2\alpha_\textrm{s}^2(2\kappa)\mathcal{A}_{-}(\omega)\mathcal{A}_{+}(\omega)\chi_\textrm{eff}(\omega)\big)^*\Big]	\\
			\phantom{S_{XY}(\omega) = }+\Im\!\Big[\big(i\hbar G_0^2\abs{\alpha_\textrm{s}}^2(2\kappa)\mathcal{A}_{-}(\omega)\mathcal{A}_{+}(\omega)\chi_\textrm{eff}(\omega)+(2\kappa)\mathcal{A}_{+}(\omega)\frac{\chi_\textrm{eff}(\omega)}{\chi_\textrm{m}(\omega)}-1\big)^*	\\
			\phantom{S_{XY}(\omega) = \Im\!\Big[}\times\big(-i\hbar G_0^2{\alpha_\textrm{s}^*}^2(2\kappa)\mathcal{A}_{-}(\omega)\mathcal{A}_{+}(\omega)\chi_\textrm{eff}(\omega)\big)\Big]	\\
			\phantom{S_{XY}(\omega) = }+2\Im\!\Big[\big(-iG_0\alpha_\textrm{s}^*\sqrt{2\kappa}\mathcal{A}_{-}(\omega)\chi_\textrm{eff}(\omega)\big)\!\times\!\big(iG_0\alpha_\textrm{s}\sqrt{2\kappa}\mathcal{A}_{+}(\omega)\chi_\textrm{eff}(\omega)\big)^*\Big]\,S_{F}^\textrm{th}(\omega)
	\end{cases}
\label{eqn: spectrum terms}
\end{equation}
\end{widetext}
The frequency-dependence of the spectral density is non-trivial, but thanks to the parametric form of Eq.~\ref{eqn: parametric spectrum} it is easy to identify the quadrature angle that minimizes it at each frequency:
\begin{equation}
	\theta_\textrm{min}(\omega) = \frac{\pi}{2}+\frac{1}{2}\arctan{\left(\frac{2S_{XY}(\omega)}{S_{X}(\omega)-S_{Y}(\omega)}\right)}.
\label{eqn: minimizing angle}
\end{equation}
In practice, unless a variational-readout setup is used~\cite{Kimble:2001:PhysRevD}, only one angle should be considered for the entire spectrum. However, we can still define $S_\textrm{min}(\omega):=S_\theta(\omega)|_{\theta=\theta_\textrm{min}(\omega)}$ as the optimal spectral density following the minimizing angle across the spectrum to obtain a more comprehensive view of optomechanical squeezing.

\begin{figure*}[]
	\centerline{\includegraphics[width=\textwidth]{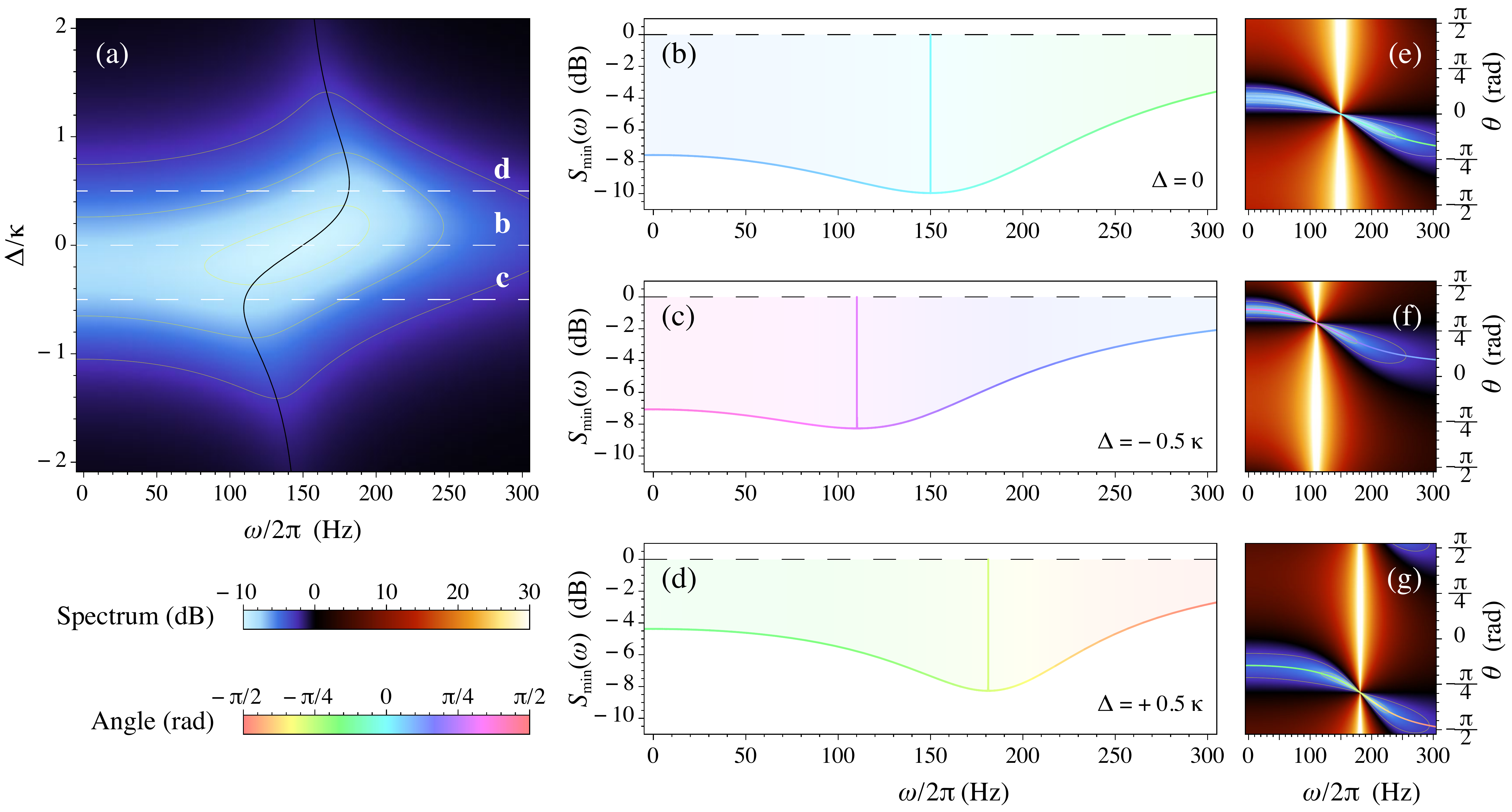}}
	\caption[]{Characterization of the spectral density of the output field noise. \;\textbf{(a)} Minimized spectrum $S_\textrm{min}(\omega)$ as a function of detuning (vertical axis). The spectrum features a singularity around the mechanical frequency, leading to no noise reduction in a narrow band whose width is inversely proportional to the quality factor of the oscillations. The narrow band is centred around the effective mechanical frequency $\omega_\textrm{eff}$, determined as a function of detuning by the optical spring effect. The mesh lines surround the regions squeezed by $\SI{3}{\deci\bel}$, $\SI{6}{\deci\bel}$, and $\SI{9}{\deci\bel}$. Dashed lines indicate the detunings chosen for parts (b-d). \;\textbf{(b-d)} Frequency-dependence of squeezing for detunings $\Delta=0$ (b), $\Delta=-0.5\kappa$ (c) and $\Delta=+0.5\kappa$ (d). For each detuning the optimal spectral density $S_\textrm{min}(\omega)$ is shown, coloured according to the quadrature angle minimizing the spectrum. The singularity at the effective mechanical frequency is seen acting as the centre of the dispersive effects leading to quadrature rotation. \;\textbf{(e-g)} Normalized spectral density $S_\theta(\omega)$ as a function of quadrature angle (vertical axis). Both squeezing (blue) and anti-squeezing (red) are more concentrated around the effective mechanical frequency, distinctly acting as a point of singularity. High precision in quadrature is required close to the dispersive centre to avoid signal contamination from the excess noise. The line in the centre of the squeezed (blue) region represents the angle minimizing the spectrum. Mesh lines showing the regions of $\SI{3}{\deci\bel}$, $\SI{6}{\deci\bel}$, and $\SI{9}{\deci\bel}$ squeezing are also displayed.}
\label{fig: spectrum}
\end{figure*}

\begin{table}[!ht]
\begin{center}
\begin{tabular}{lccc}
	\textbf{Parameter}		&	\multicolumn{2}{l}{\textbf{Symbol}}	&	\textbf{Value}					\\\hline\hline
						&\phantom{\quad}&					&								\\
	Mechanical frequency	&&	$\omega_\textrm{m}$			&	$2\pi\times\SI{150}{\hertz}$		\\
	Mechanical quality factor	&&	$Q_\textrm{m}$					&	$5\times10^6$					\\
	Oscillator mass			&&	$m$							&	$\SI{0.5}{\kilo\gram}$			\\
	Temperature			&&	$T$							&	$\SI{3}{\milli\kelvin}$				\\
						&&								&								\\
	Input power			&&	$P_\textrm{in}$					&	$\SI{20}{\watt}$					\\
	Wavelength			&&	$\lambda$						&	$\SI{1064}{\nano\metre}$			\\
	Free spectral range		&&	$\omega_\textrm{FSR}$			&	$2\pi\times\SI{1}{\giga\hertz}$		\\
	Cavity damping			&&	$\kappa$						&	$2\pi\times\SI{0.5}{\mega\hertz}$	\\
	Finesse				&&	$\mathcal{F}$					&	$1000$						\\
	Reduced OM coupling 	&&	$g_0$						&	$2\pi\times\SI{0.63}{\milli\hertz}$	\\
						&&								&								\\
	Test masses (interf.)		&&	$m_\textrm{gw}$				&	$\SI{40}{\kilo\gram}$				\\
	Cavity length (interf.)		&&	$L_\textrm{gw}$				&	$\SI{4}{\kilo\metre}$				\\
	Cavity damping (interf.)	&&	$\kappa_\textrm{gw}$			&	$2\pi\times\SI{100}{\hertz}$		\\
	Operating power (interf.)	&&	$P_\textrm{gw}$				&	$\SI{10.6}{\kilo\watt}$			\\
	\\\hline
\end{tabular}
\end{center}
\caption{Parameters characterizing the optomechanical system, used for the generation of squeezed light, and the gravitational-wave interferometer, where the squeezed light is injected. The reported values are those used to obtain the results in Fig.~\ref{fig: spectrum} and~\ref{fig: sensitivity}.}
\label{tab: parameters}
\end{table}

\section{Squeezed spectrum}
\label{sec: Squeezed spectrum}

With all the key elements at our disposal, we can now examine a few different cases in the frequency band of interest. Even though a higher intra-cavity power would facilitate the optomechanical interaction, we are going to consider a cavity with medium finesse and short lifetime. The reason behind this choice is to have the squeezing disperse over different quadratures purely because of optomechanics, and not because of the filtering action of the cavity. Moreover, the parameter requirements for the oscillator will be chosen to have a cap on the squeeze factor of $\SI{10}{\decibel}$, corresponding to a noise level of $0.1$, to allow realistic comparison with a traditional squeezing source with the same performance. The moving mirror considered has a mass $m=\SI{0.5}{\kilo\gram}$, mechanical frequency $\omega_\textrm{m}=2\pi\times\SI{150}{\hertz}$ and a quality factor $Q_\textrm{m} = 5\times10^6$ at a cryogenic temperature of $T=\SI{3}{\milli\kelvin}$. The cavity resonance is tuned to a wavelength $\lambda=\SI{1064}{\nano\metre}$, with free spectral range $\omega_\textrm{FSR}=\pi c/L=2\pi\times\SI{1}{\giga\hertz}$ and half-linewidth $\kappa=2\pi\times\SI{0.5}{\mega\hertz}$. The input power $P_\textrm{in}=\hbar\omega_\textrm{o}\abs{\alpha_\textrm{in}}^2=\SI{20}{\watt}$ is set to conform to the operational requirement of the new generation of interferometers. The reduced optomechanical coupling, dependent on the zero-point fluctuation $x_\textrm{ZPF}=\sqrt{\hbar/(2m\omega_\textrm{m})}$, is $g_0:=G_0x_\textrm{ZPF}=2\pi\times\SI{0.63}{\milli\hertz}$. These values are listed in Table~\ref{tab: parameters} for convenience.

\begin{figure*}[]
	\centerline{\includegraphics[width=\textwidth]{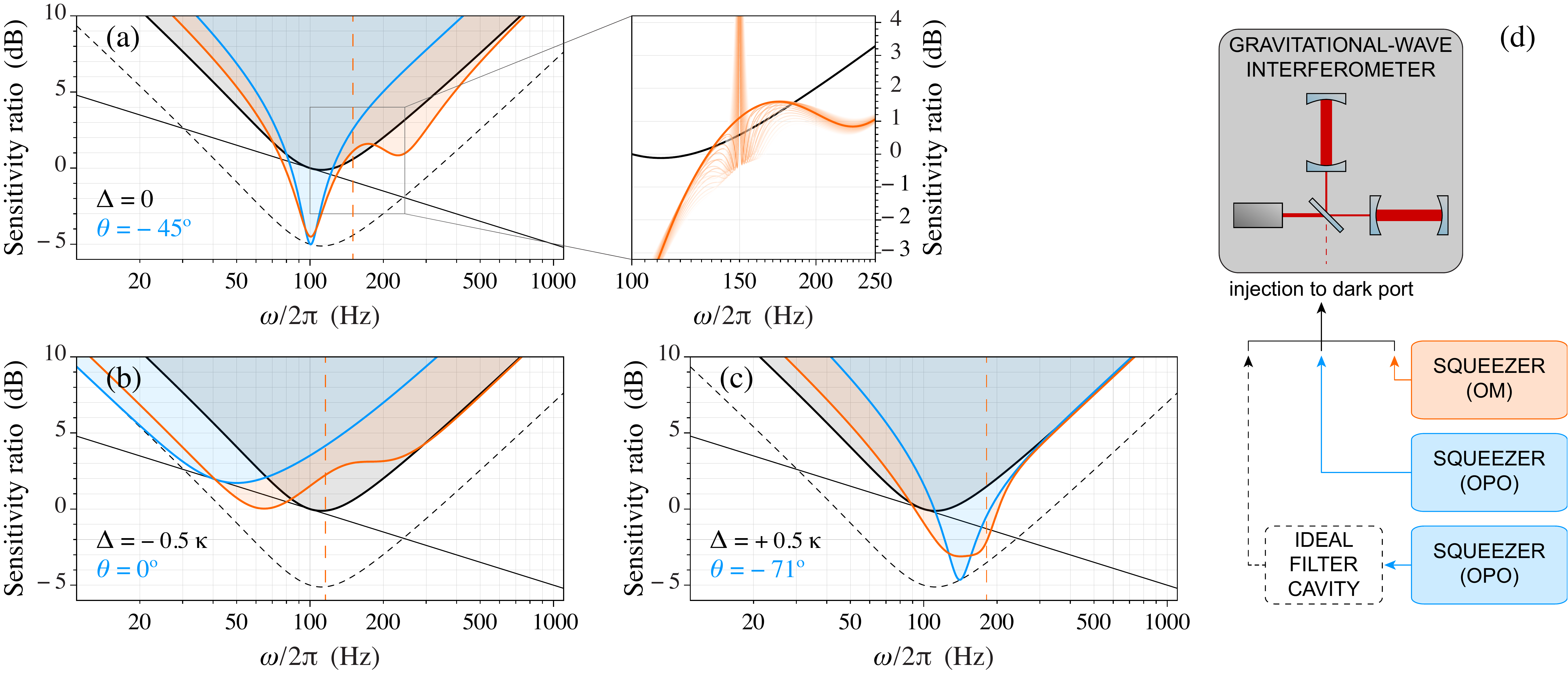}}
	\caption[]{\;\textbf{(a-c)} Sensitivity of LIGO operating at $P_\textrm{gw}=P_\textrm{SQL}$ to reach the standard quantum limit. The comparison is performed for three different detunings: $\Delta=0$ (a), $\Delta=-0.5\kappa$ (b) and $\Delta=+0.5\kappa$ (c). All traces are obtained from the square root of the corresponding spectral densities, normalized by $\omega/\kappa_\textrm{gw}$ and by the SQL $S_\textrm{SQL}(\omega)$ (straight black line) calculated at $\omega=\kappa_\textrm{gw}$. The sensitivity of a conventional interferometer obtained from $S_\textrm{gw}(\omega)$ (solid black line) is limited, in conditions without other technical noises, by the SQL. When $\SI{10}{dB}$ of squeezing with ideal quadrature rotation are injected into the interferometer, the sensitivity obtained from $\bar{S}_\textrm{gw}(\omega)|_{\theta=\varphi(\omega)}$ (dashed black line) has a noise floor lowered across the full spectrum. The spectral density $\bar{S}_\textrm{gw}(\omega)$ for a fixed quadrature (blue trace) gives a similar enhancement in sensitivity only on a limited region, determined by the quadrature being squeezed: amplitude squeezing has noise reduction in the low-frequency end of the spectrum, phase-squeezing in the higher end, and hybrid squeezing on a narrow band in between. The sensitivity of an interferometer injected with optomechanical squeezing (orange trace) results from $\bar{S}_\textrm{gw}(\omega)|_{\theta=\theta_\textrm{min}(\omega)}$, and features a modified enhancement thanks the frequency dependence of both the squeezing level and the squeezed angle. The zoomed-in inset of (a) shows how the singularity around the mechanical frequency $\omega_\textrm{m}$ (dashed orange line) affects the sensitivity: starting from the original trace corresponding to the minimizing angle, subsequent traces (light orange to white) are obtained by accounting for a total deviation of $\SI{6}{\percent}$ of a radian from $\theta_\textrm{min}(\omega)$ in intervals of $\SI{0.3}{\percent}$. A similar behaviour should be expected in (b) and (c), although in both cases the detuning causes the effective mechanical frequency to be shifted at a different value. \textbf{(d)} A schematic diagram showing how squeezing may be injected in a gravitational-wave detector. The squeezed source enters the dark port of the interferometer from an optomechanical (OM) system (orange), from a traditional squeezer (blue) or from a traditional squeezer after passing through a filter cavity to implement ideal quadrature rotation (dashed black).}
\label{fig: sensitivity}
\end{figure*}

The system defined by this parameter selection attains squeezing over different quadratures in a band of a few hundred $\SI{}{\hertz}$, as shown in Fig.~\ref{fig: spectrum}. This is what is required for a comparison with traditional fixed-quadrature squeezing in connection with a gravitational-wave interferometer, whose best sensitivity is determined by its operating power $P_\textrm{gw}$ and usually corresponds to $\sim\SI{100}{\hertz}$. The strongest dispersive effects take place close to the effective mechanical frequency $\omega_\textrm{eff}:=1/\sqrt{\Re(m\chi_\textrm{eff})}$, which acts as a focal point for the coupled dynamics of the moving mirror and the optical field. The optimal spectrum $S_\textrm{min}(\omega)$ culminates at frequencies $\omega=\omega_\textrm{eff}$ into a peak-like feature, due to a singular inversion of the frequency response of the system, and the measurement here can at best match the original shot noise. This is featured in Fig.\ref{fig: spectrum}a as a black line, varying with detuning according to the optical spring effect. As Fig.~\ref{fig: spectrum}e-g show, a small deviation from the quadrature $\theta_\textrm{min}$ can lead to significant excess noise in the region close to $\omega_\textrm{eff}$, due to the strong anti-squeezing concentrated around this point. The width of this effect is inversely proportional to $Q_\textrm{m}$; even a quality factor of $50$, five orders of magnitude lower than the one considered, would not extend the linewidth of the dispersive peak over $\SI{1}{\hertz}$. Moreover, the possibility of changing the detuning allows control over what part of the spectrum would be most influenced.

Limiting the observations to a region of $\SI{3}{\deci\bel}$ of squeezing around the dispersive centre, one can see that at $\Delta=0$ (Fig.~\ref{fig: spectrum}b and~\ref{fig: spectrum}e) the squeezed angle varies from about $\pi/12$ at DC to about $-\pi/6$ at $\SI{300}{\hertz}$, achieving rotation by $\pi/4$ overall. Slightly larger rotations are obtained at a detuning of $\Delta=-0.5\kappa$ (Fig.~\ref{fig: spectrum}c and~\ref{fig: spectrum}f) and $\Delta=+0.5\kappa$ (Fig.~\ref{fig: spectrum}d and~\ref{fig: spectrum}g). Only at high mechanical frequencies would one obtain a full $\pi/2$ rotation across the entire spectrum, but it should also be considered that far from $\omega_\textrm{eff}$ the interaction is not strong enough to correlate the noise of different quadratures and the squeezing is much more diluted. High detunings, too, see a rotation asymptotically close to $\pi/2$, but again noise reduction becomes negligible and there is close to no advantage for $\abs{\Delta}\gtrsim1.5\kappa$.

\section{LIGO sensitivity enhancement}
\label{sec: LIGO sensitivity enhancement}

In interferometry, back action leads to a fundamental limit to the measurement capability known as the standard quantum limit. The SQL is a consequence of shot noise, which introduces fluctuations in the output field detected, and of radiation pressure noise, with light pushing the test masses into a random motion that could mask the measurement. The SQL has spectral density~\cite{Kimble:2001:PhysRevD}
\begin{equation}
	S_\textrm{SQL}(\omega)=\frac{8\hbar}{L_\textrm{gw}^2m_\textrm{gw}\omega^2},
\label{eqn: standard quantum limit}
\end{equation}
where $L_\textrm{gw}$ is the arms' length and $m_\textrm{gw}$ is the mass of the test mirrors. The noise floor determined by the SQL imposes an intrinsic bound to the sensitivity of the interferometer, and the quantity
\begin{equation}
	P_\textrm{SQL}:=\frac{L_\textrm{gw}^2m_\textrm{gw}\kappa_\textrm{gw}^4}{2\omega_\textrm{o}}
\label{eqn: SQL power}
\end{equation}
represents the reference power required to achieve best sensitivity in the band determined by the linewidth of the arm cavities, $\kappa_\textrm{gw}$. Table~\ref{tab: parameters} reports the values of $m_\textrm{gw}$, $L_\textrm{gw}$, $\kappa_\textrm{gw}$ and $P_\textrm{SQL}$ in use in the advanced LIGO interferometer. The sensitivity of the interferometer is dictated by its minimum noise spectrum, and for a standard interferometer operating at $P_\textrm{gw}=P_\textrm{SQL}$ this is
\begin{equation}
	S_\textrm{gw}(\omega)=\frac{4\kappa_\textrm{gw}^8+\omega^4(\kappa_\textrm{gw}^2+\omega^2)^2}{4\kappa_\textrm{gw}^4\omega^2(\kappa_\textrm{gw}^2+\omega^2)}S_\textrm{SQL}(\omega).
\label{eqn: standard sensitivity}
\end{equation}
The contribution of shot noise is limited at frequencies lower than $\kappa_\textrm{gw}$ thanks to the cavity-enhanced reading, and becomes predominant only at higher frequencies. Radiation pressure noise is tied to the mechanical susceptibility of the test masses and its effects are, in contrast, limited to the low-frequency region of the spectrum. So far, the lower spectral region has however been dominated by other technical noises, such as the thermal noise of the optical coatings. The development of new technologies~\cite{Cole:2013:NatPhot} has allowed significant breakthroughs, allowing the new generations of gravitational-wave detectors to push the sensitivity at low frequencies to the fundamental limit imposed by radiation pressure noise.

Refined readings are possible with the injection of squeezed light, pushing the capability of the interferometer beyond the SQL. Injecting light squeezed by a factor $e^{-2r}$ at an angle $\theta$ modifies the spectral density to
\begin{equation}
	\bar{S}_\textrm{gw}(\omega) = S_\textrm{gw}(\omega)\Big[\cosh{(2r)}-\sinh{(2r)}\cos{(2\theta-2\varphi(\omega))}\Big],
\label{eqn: squeezed-input sensitivity}
\end{equation}
where $\varphi(\omega):=-\arccot{\left(\frac{2\kappa_\textrm{gw}^4}{\omega^2(\kappa_\textrm{gw}^2+\omega^2)}\right)}$ is the element of angle rotation due to back action of the test masses. It is for ideal squeezing rotation, i.e.\ $\theta(\omega)=\varphi(\omega)$, that the spectral density experiences a global reduction equal to the squeezing factor $e^{-2r}$. If the squeezing angle $\theta$ is fixed and does not rotate over frequencies, the same reduction will occur only over a reduced band. When phase squeezing ($\theta=-\pi/2$) is injected, the photon-counting noise is effectively reduced, improving sensitivity at higher frequencies. However, the motion of the test masses becomes subject to higher fluctuations induced by anti-squeezing on the amplitude quadrature, and the sensitivity at lower frequencies is compromised. The impact of phase squeezing is equivalent to having higher power available in the arms of the interferometer, without the complications arising from increased thermal noise on the test masses and their suspension. Squeezing of the amplitude quadrature ($\theta=0$) prompts the opposite effect, and sensitivity is enhanced in the lower-frequency band while noise increases at the other side of the spectrum. Hybrid squeezing ($-\pi/2<\theta<0$), can beat the SQL close to the frequency of optimal sensitivity for gravitational waves at $\SI{100}{\hertz}$, but the increased noise due to traces of anti-squeezing in both phase and amplitude degrade the quality of the measurement outside of this bandwidth. This kind of measurement could be useful for detection enhancement of signals from known gravitational wave sources, like pulsars.

As we have shown, squeezing obtained via an optomechanical system has an inherent frequency dependence that could compensate for the effects of $\varphi(\omega)$ over a wider section of the measurement band compared to squeezing at a fixed quadrature. However, in the optomechanical system it is not just the squeezed angle that is frequency dependent: the squeezing factor, as inferred from Fig.~\ref{fig: spectrum}b-d, also varies across the spectrum. This seemingly undesirable property could prove advantageous if we consider that the frequency dependence of the squeezed quadrature might not accomplish the full rotation desired. When the squeezed quadrature deviates significantly from $\varphi(\omega)$, the measurement would mostly be obstructed by anti-squeezing and it would be preferable to renounce all cross-correlations between the quadratures. This is also common to fixed-quadrature squeezing, where the sensitivity is degraded by anti-squeezing in half of the total spectrum. Thus, even if broadband enhancement is not achieved, the absence of interaction away from the dispersive centre could be used to one's advantage if the system is prepared to provide high squeezing in a region with reduced quadrature rotation (for example, from $-\pi/2$ to $-\pi/4$) and no change from a conventional interferometer elsewhere.

The performance of gravitational-wave interferometers in different set-ups is shown in Fig.~\ref{fig: sensitivity}, which compares the sensitivity of a traditional interferometer with no squeezing injection (black) to that obtained by injecting frequency-independent squeezing from a traditional source (blue) and frequency-dependent squeezing obtained via optomechanics (orange). The sensitivity traces are obtained as the square root of each corresponding spectral density, and they are all normalized by the SQL of Eq.~\ref{eqn: standard quantum limit} calculated at $\omega=\kappa_\textrm{gw}$ and additionally rescaled in terms of frequency by $\omega/\kappa_\textrm{gw}$. The squeezing at a fixed quadrature is taken with a uniform squeeze factor of $\SI{10}{\deci\bel}$, comparable to the finest optomechanical squeezing achievable with the chosen parameters. Realistic interferometers are subject to propagating losses that may reduce the operative squeezing from $\SI{10}{\deci\bel}$ to $\SI{6}{\deci\bel}$~\cite{Oelker:2014:OptExp}, but as this would apply equally regardless of the source we consider the former for a clearer comparison. At no detuning ($\Delta=0$, Fig.~\ref{fig: sensitivity}a), optomechanical squeezing can perform generally better than fixed hybrid squeezing at all frequencies; the advantage is particularly noticeable at frequencies higher than $\SI{100}{\hertz}$, where most of the rotation takes place and the difference between the two traces gets as high as $\SI{5.5}{\deci\bel}$ before the interaction becomes too weak and the trace converges to that of a conventional interferometer. The situation is similar with a negative detuning ($\Delta=+0.5\kappa$, Fig.~\ref{fig: sensitivity}b): compared with pure amplitude squeezing, optomechanical squeezing avoids the additional noise introduced by anti-squeezing at high frequencies while still achieving noise cancellation in the lower end of the spectrum, although of slightly lower calibre. A positive detuning ($\Delta=+0.5\kappa$, Fig.~\ref{fig: sensitivity}c) can result in a broad sensitivity enhancement in the region between $\SI{100}{\hertz}$ and $\SI{200}{\hertz}$ without the need to sacrifice too much sensitivity at lower frequencies, as opposed to the frequency-independent squeezing that can not neutralize anti-squeezing in the wrong quadrature. One should remember that there is a singularity at the effective mechanical frequency $\omega_\textrm{eff}$, and deviation from the optimal quadrature close to this frequency might introduce undesired noise into the system. Considering imperfections in the precision of the lock systems, we can see the effects of deviations by up to $\SI{6}{\percent}$ of a radian on the sensitivity in the inset of Fig.~\ref{fig: sensitivity}a. Quadratures different from $\theta_\textrm{min}(\omega)$ induce a small region of better sensitivity before a spike of overwhelming noise takes over around the resonance. As the angle gets closer to the optimal one, this effect spans a narrower region until it gets completely cancelled for $\theta=\theta_\textrm{min}(\omega_\textrm{eff})$, for which the spectrum is identical to the original shot noise. The reason why better sensitivity is still possible for quadratures that differ from $\theta_\textrm{min}(\omega)$ lies in the fact that the rotation achieved by the optomechanical system is not the ideal one required by the gravitational wave detector, and a slight deviation may actually bring it closer.

\section{Conclusions}
\label{sec: Conclusions}

We characterized the frequency-dependent properties of optomechanical squeezing and found that the quadrature rotation observed could have interesting applications in gravitational-wave detectors. Compared to squeezing from a traditional OPO source, injection of optomechanically-generated squeezing can be both beneficial and detrimental for the sensitivity of the interferometric signal. The inherent quadrature rotation extracted from the coupled dynamics of the light and the resonator is an attractive option to counteract the combined action of radiation pressure and shot noise, and could provide an elegant alternative to rotation obtained through a filter cavity. We project a sensitivity enhancement of up to $\SI{5.5}{\deci\bel}$ in the measurement band above $\SI{100}{\hertz}$ when the two methods are compared at a cap of maximum squeezing of $\SI{10}{\deci\bel}$. Additionally, the response of the optomechanical cavity to different detunings offers extended flexibility and a broader choice on the type of enhancement accessible. The extensive efforts placed by the gravitational-wave community in the suppression of mechanical noise, particularly in the bandwidth of interest, should pose a strong foundation for the experimental implementation of an optomechanical system with specifications similar to the ones suggested. Unfortunately there could be several obstacles to be overcome before the realization of a full implementation: for example the thermal requirements might prove hard to meet, or locking might not be good enough and oppressive noise due to anti-squeezing of nearby quadratures could compromise the measurement. On the other hand, the technical feasibility of cavity-induced rotation at $\SI{50}{\hertz}$ has yet to be demonstrated. A cavity approaching that required exists~\cite{Della-Valle:2014:OptExp}, but the losses are still unfavourable.

\section{Acknowledgements}
\label{sec: Acknowledgements}
This research was funded by the Australian Research Council Centre of Excellences CE110001027, the Discovery Project DP150101035. PKL is supported by the ARC Laureate Fellowship FL150100019, BCB by the ARC future fellowship FT100100048.

%\bibliographystyle{plainbib}
%\bibliography{papersLIGO}


\end{document}